\documentclass[prl,twocolumn,showpacs,floatfix,amsmath,amssymb,superscriptaddress]{revtex4}

\usepackage{graphicx}% Include figure files

\usepackage{dcolumn}% Align table columns on decimal point

\usepackage{bm}% bold math

\usepackage{subfigure}

\usepackage{amssymb}
\usepackage{hyperref}

\usepackage{array}
\usepackage{amsthm}
\usepackage{amsmath}
\usepackage{epsfig}
\usepackage{graphicx}
\usepackage{wasysym}
\usepackage{bbm}
\usepackage[all]{xy}

\def \be{\begin{equation*}}
\def \ee{\end{equation*}}

\begin{document}

\title{Fermionic Hopf solitons and Berry's phase in topological surface superconductors}

\author{Ying Ran}
\altaffiliation[Present address~: ]{Department of Physics,
Boston College, Chestnut Hill, MA 02467, USA} \affiliation{Department of Physics,
University of California, Berkeley, CA 94720, USA}\affiliation{Materials Sciences Division,
LBNL Berkeley, CA 94720, USA}
\author{Pavan Hosur}
\affiliation{Department of Physics,
University of California, Berkeley, CA 94720, USA}
\author{Ashvin Vishwanath}
\affiliation{Department of Physics, University of California,
Berkeley, CA 94720, USA} \affiliation{Materials Sciences Division,
LBNL Berkeley, CA 94720, USA}

\begin{abstract}
A fascinating idea in many body physics is that quantum statistics
may be an emergent property.
This was first noted in the Skyrme model of nuclear matter, where a
theory formulated entirely in terms of a bosonic order parameter
field contains fermionic excitations. These excitations are smooth
field textures, and believed to describe neutrons and protons.
We argue that a similar phenomenon occurs in topological insulators
when superconductivity gaps out their surface states. Here, a smooth
texture is naturally described by a three component real vector. Two
components describe superconductivity, while the third captures the
band topology.
Such a vector field can assume a 'knotted' configuration in three
dimensional space - the Hopf texture - that cannot smoothly be
unwound. Here we show that the Hopf texture is a fermion. To
describe the resulting state, the regular Landau-Ginzburg theory of
superconductivity must be augmented by a topological Berry phase
term. When the Hopf texture is the cheapest fermionic excitation,
interesting consequences for tunneling experiments are predicted.

\end{abstract}

\maketitle

There has been much recent excitement relating to topological
insulators (TIs), a new phase of matter with protected surface
states \cite{hasan_kane_review,qi:33}. Particularly rich phenomena
are predicted to arise when this phase is combined with conventional
orders such as magnetism
\cite{PhysRevLett.102.146805,anomalous_hall_TI}, crystalline
order\cite{disloc_nature} and superconductivity. The last is
particularly interesting. Superconductivity induced on the surface
of a TI was predicted to have vortices harboring Majorana zero modes
\cite{fu:096407}, similar to the bound states in vortices of a $p+ip$ superconductor\cite{Jackiw1981681,PhysRevB.61.10267,PhysRevLett.86.268}. These are of interest to quantum information
processing, since they are intrinsically robust against errors.
Recently, superconductivity was discovered in a doped TI
\cite{PhysRevLett.104.057001}, which could be used to induce surface
superconductivity. Below we discuss a new theoretical approach to
studying this remarkable superconducting phase, which provides
different insights and directions for experiments.

We focus on smooth configurations where the energy gap never
vanishes. In this case, the low energy description of the system is
entirely in terms of bosonic 'order parameter' coordinates, much as
the Landau Ginzburg order parameter theory describes superconductors
at energies below the gap. Can fermions ever emerge is such a
theory? While it is easy to imagine obtaining bosons from a
fermionic theory, the reverse appears impossible at first sight.
However, it has been shown in principle that bosonic theories that
contain additional Berry's phase (or Wess-Zumino-Witten) terms, can
accomplish this transmutation of statistics. We show that this
indeed occurs in the superconductor-TI (Sc-TI)system; the order
parameter theory contains a Berry phase term which implies that a
particular configuration of fields - the Hopf soliton (or {\em
Hopfion}) - carries fermionic statistics. While such statistics
transmutation is common in one dimension eg. bosons with hardcore
interactions \cite{Giamarchi}, it is a rare phenomenon in higher
dimensions. In the condensed matter context, an physically
realizable  example exists in two dimensions: solitons of quantum
Hall ferromagnets (skyrmions) are fermionic and charged, and have
been observed\cite{PhysRevB.47.16419,PhysRevLett.64.1313,PhysRevLett.74.5112}. However,
the superconductor-TI system is, to our knowledge, the first
explicit condensed matter realization of this phenomenon in three
dimensions.

\section{Model and Hopf Texture}
The essential properties of a topological insulator are captured by
a simplified low energy theory with a three dimensional Dirac
dispersion (a microscopic realization is described later): $H_D =
\psi^\dagger [v_F{\bf \alpha}\cdot {\bf p}+m\beta_0]\psi$, where
$(\alpha_1,\,\alpha_2,\,\alpha_3,\,\beta_0)$ are $4\times4$
anti-commuting matrices which involve both spin and sublattice
degrees of freedom. The matrices $\alpha_i$ can be taken to be
symmetric, while $\beta_{0}$ and
$\beta_5=\alpha_1\alpha_2\alpha_3\beta_0$ are antisymmetric. The
dispersion then is $\epsilon(p)=\pm \sqrt{v_F^2p^2+m^2}$. An
insulator is obtained for $m\neq0$. Changing the sign of $m$ results
in going from a trivial to a topological insulator. Which sign of
'm' is topological is set by the band structure far away from the
node - here we assume $m<0$ is topological. Since the vacuum can be
taken to be a trivial insulator, this mass term changes sign at the
topological insulator surface. Consider now adding (onsite)
superconducting pairing, which may be proximity induced by an s-wave
superconductor. Then
 $H_{\rm pair} = \Delta \psi^\dagger \beta_5\psi^\dagger+{\rm
 h.c.}$, and we can write the total Hamiltonian $H_f=H_D+H_{\rm pair}$
 as:
\begin{equation}
H_f = \left [\begin{array}{cc}
      \psi^\dagger & \psi
    \end{array} \right ]
    \left [ \begin{array}{cc}
    -iv_F{\bf \alpha}\cdot {\bf \partial}+m\beta_0 & \Delta \beta_5 \\
      \Delta^* \beta_5 & -iv_F{\bf \alpha}\cdot {\bf \partial}+m\beta_0
    \end{array} \right ]
    \left [ \begin{array}{c}
              \psi \\
              \psi^\dagger
            \end{array}
    \right ]
\label{HMF}
\end{equation}

the spectrum now is $\epsilon(p)=\pm \sqrt{v_F^2p^2+M^2}$ where $M^2
= m^2+|\Delta|^2$. It is convenient to define a three vector
$\vec{n} = ({\mathcal Re}\Delta,{\mathcal Im}\Delta, m)$, such that
$|\vec{n}|=M$. A singular configuration is one where all three
components of this vector go to zero. This corresponds to the core
of a vortex (where the components of the pairing $\Delta$ vanish),
intersecting the surface of a topological insulator (where the third
component is zero). It has been pointed out that an odd strength
vortex will give rise to an unpaired Majorana zero mode in this
configuration \cite{fu:096407}, which can also be viewed as a
hedgehog \cite{PhysRevLett.104.046401} with odd integer topological
charge. Note, at the core of these singular configurations the gap
closes, allowing for the possibility of localized bound states near
zero energy. In this work we will {\em only} consider smooth
textures of the $\vec{n}$ field, where the single particle gap is
nonzero everywhere. An effective theory of slow fluctuations of the
'order parameter' field $\vec{n}(r,t)$ (occuring over spatial (time)
scales much larger than $\xi=\hbar v_F/M$ ($\tau=\xi/v_F$)), can be
obtained by integrating out the gapped fermions. An analogous
procedure is well know in the context of the BCS theory of
superconductivity, where it leads to the Landau-Ginzburg action.
Here we will find that an extra topological term arises, that
transmutes statistics and leads to fermionic solitons.

\begin{figure}
\includegraphics[width=0.4\textwidth]{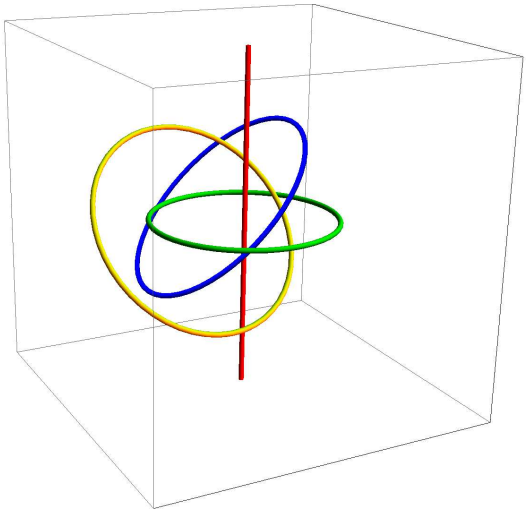}
\caption{The Hopf map: $f_H:\vec{r}\rightarrow \hat{n}$ is shown by
displaying contours of equal $\hat{n}$. Points at infinity are all
mapped to the same point on the sphere $f_H(\infty)=\hat{z}$. In red
is $f^{-1}_H[\hat{n}=\hat{z}]$, in green  $f^{-1}_H[-\hat{z}]$, in
blue $f^{-1}_H[\hat{x}]$, and in yellow $f^{-1}_H[\hat{y}]$. Note
the unit linking of any pair of curves, which can be used to define
the Hopf texture.}
 \label{fig1}
\end{figure}

Consider a smooth configuration of $\vec{n}(r)$, which  can be
normalized to give a unit vector $\hat{n}(r)$ at each point. This
defines a mapping from each point of three dimensional space, to a
unit three vector, which describes the surface of a sphere $S^2$. We
require that the mapping approaches a constant at infinity:
$\hat{n}(|r|\rightarrow \infty)={\rm const}$ (e.g the vacuum). Can
all such mappings be smoothly distorted into one another? A
surprising result due to Hopf \cite{Hopf} 1931, is that there are
topologically distinct mappings, which can be labeled by distinct
integers $h$ (the Hopf index). No smooth deformation can connect
configurations with different Hopf indices. Mathematically, Hopf
showed that the homotopy group: $\Pi_3[S^2]=Z$. A straightforward
way to establish the index is to consider the set of points in space
that map to a particular orientation of $\hat{n}$. In general this
is a curve. If we consider two such orientations, we get a pair of
curves. The linking number of the curves is the Hopf index. A
configuration with unit Hopf index can be constructed by picking a
reference vector $\hat{n}=\hat{z}$, say, and rotating it by the
following set of rotations. Any rotation is parameterized by an
angle and a direction of rotation. If we take the angle to vary as
we move in the radial direction, from $\theta=0$ at the origin, to
$\theta=2\pi$ at radial infinity, and take the axis of rotation to
be the radial direction $\hat{r}$, this gives a Hopf texture. Using
an SU(2) matrix $U(\vec{r})=
e^{i\frac{\theta(r)}{2}\hat{r}\cdot\sigma}$ to represent this
rotation, the vector field $\hat{n}(\vec{r}) = U(\vec{r})\sigma_z
U^\dagger(\vec{r})$ is the unit Hopf texture. This is readily
verified by studying which spatial points map to
$\hat{n}=\pm\hat{z}$. While the former includes all points at
infinity as well as the $z$ axis, the latter is a circle in the $xy$
plane. Clearly these two curves have unit linking number unity (see
Figure \ref{fig1}. What is the physical interpretation of this Hopf
texture in the context of TIs? A torus of TI (Figure \ref{fig2}) has
superconductivity induced on its surface. There is vacuum far away
and through the hole of the torus, which counts as a trivial
insulator, $\hat{n}=\hat{z}$. The center of the strong topological
insulator corresponds to $\hat{n}=-\hat{z}$. On the topological
insulator surface $n_z=0$, and the superconducting phase varies such
that there is a unit vortex trapped in each cycle of the torus. We
now argue that such a texture is a fermion.

\section{Hopf Solitons are Fermions}
The ground state of the mean field Hamiltonian (\ref{HMF}) for a
general texture,  has rather low symmetry, and cannot be labeled by
spin or U(1) charge quantum numbers. The only quantum number that
can be assigned is the parity of the total number of fermions
$(-1)^F$. Superconducting pairing only changes the number of
fermions by an even number, hence one can assign this $Z_2$ fermion
parity quantum number to any eigenstate. We now argue that the
fermion parity of a smooth texture is simply the parity of its Hopf
index.

First, we argue that the ground state with a topologically trivial
texture has an even number of fermions. As a representative,
consider a configuration where the superconductor pairing amplitude
is real. This is a time reversal invariant Hamiltonian. If the
ground state had an odd number of fermions, it must be doubly
degenerate at least, by Kramers theorem. However, the ground state
of any smooth texture is fully gapped and hence unique. Thus, this
configuration must have an even  fermion parity. Now, any other
texture in the same topological class can be reached by a continuous
deformation, during which the gap stays open. The fermion parity
stays fixed during this process. Note, this argument cannot be
applied to configurations with nonzero Hopf index, since these
necessarily break time reversal symmetry. For example, the
configuration shown in Figure 1, contains a pair of vortices.

To find the fermion parity of the nontrivial Hopf configurations, we
consider evolving the Hamiltonian between the trivial and $h=1$
configuration. In this process we must have $\vec{n}=0$ at some
point, which will allow for the gap to close, and a transfer of
fermion parity to potentially occur. Indeed, as shown below in
separate calculations, a change in fermion parity is induced when
the Hopf index changes by one.

\begin{figure}
\includegraphics[width=0.4\textwidth]{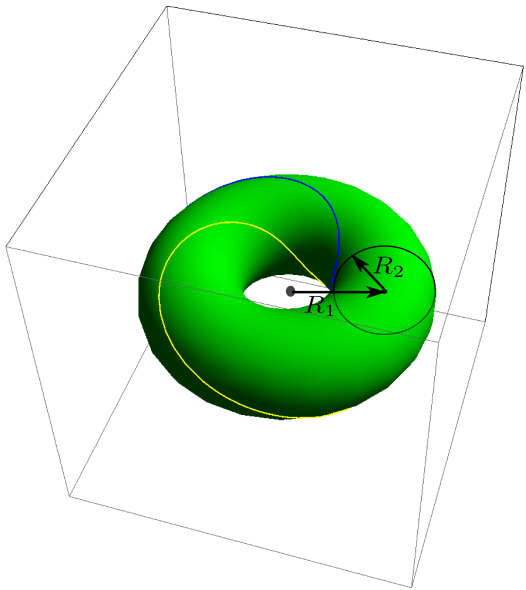}
\caption{This figure shows a torus of strong topological insulator
(green) in vacuum, whose surface is superconducting . There is unit
superconductor phase winding about each cycle of the torus. We plot
the equal phase contours on the surface whose pairing phase is
$0$(blue) and $\pi$(yellow). The unit linking of these curves
indicates this is the Hopf mapping.} \label{fig2}
\end{figure}
\begin{figure}
\begin{center}
 \includegraphics[width=0.4\textwidth]{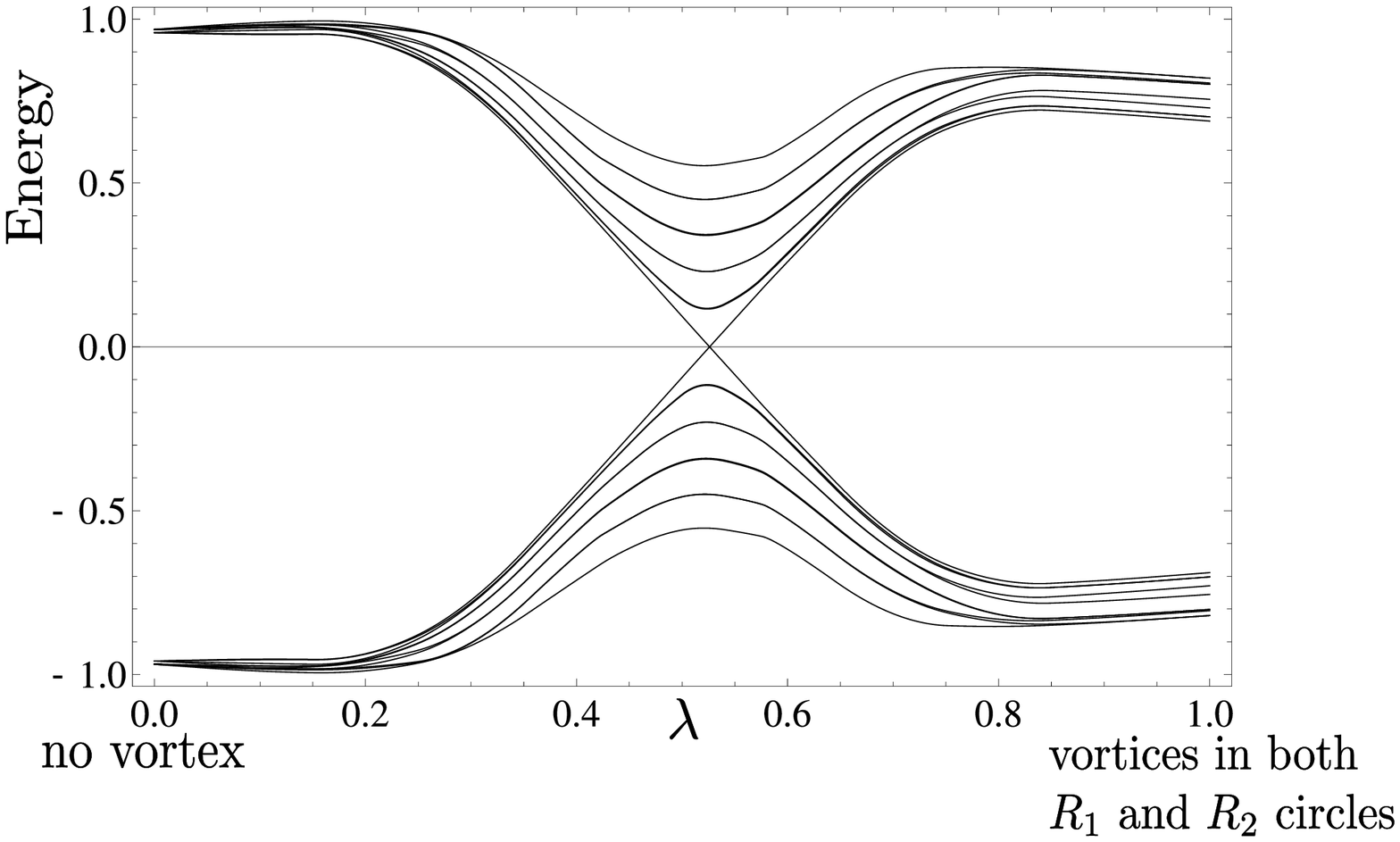}\\
\includegraphics[width=0.4\textwidth]{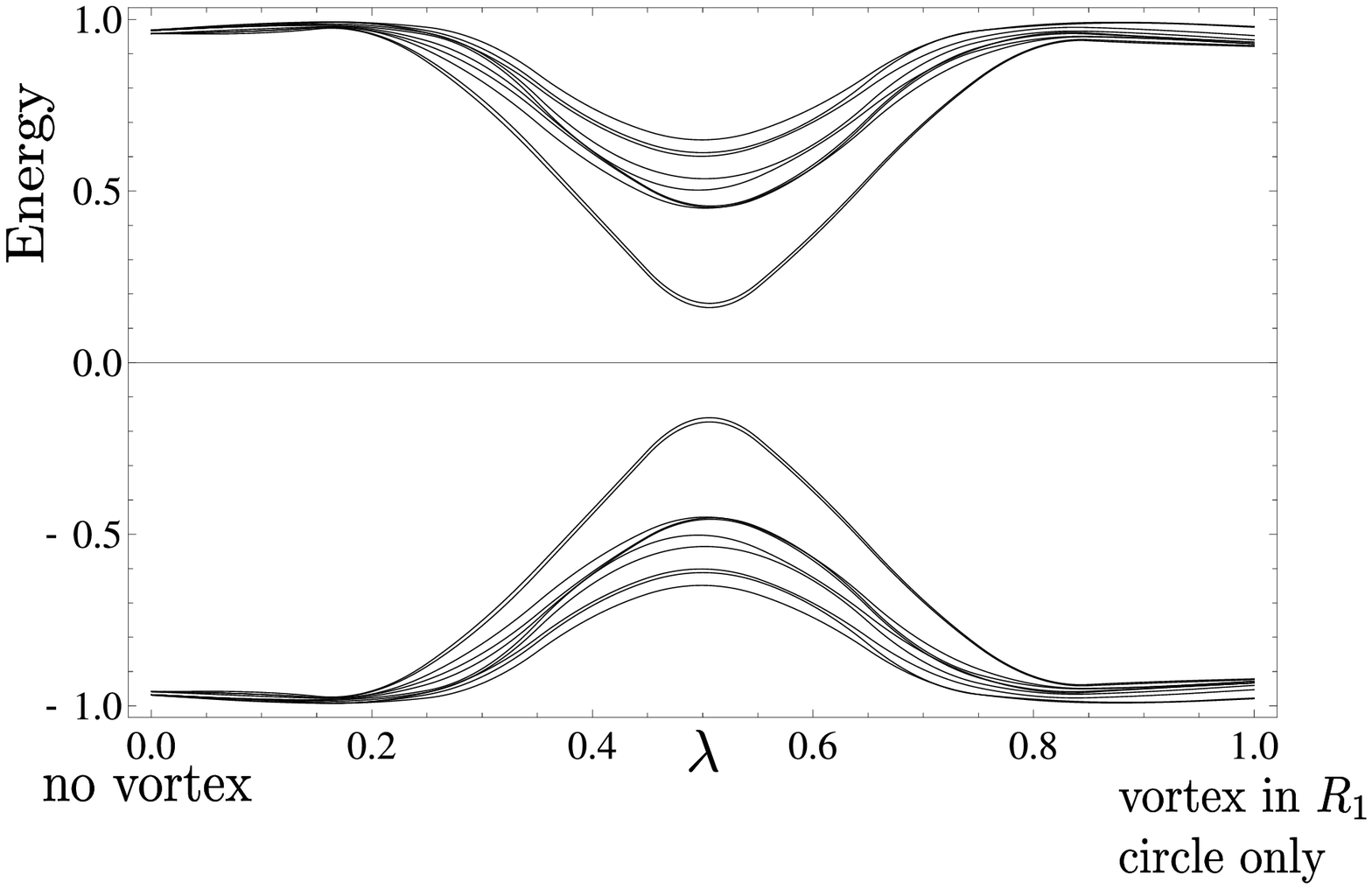}\\
\caption{The spectral flow of the lowest 20 eigenvalues when the
pairing on the surface of the topological insulator in Fig.
\ref{fig2} is linearly interpolated between two limits:
$\Delta(x,\lambda)=(1-\lambda)\Delta_0(x)+\lambda\Delta_1(x)$.
$\Delta_0(x)$ is constant over the whole surface. TOP: $\Delta_1(x)$
has a unit phase winding (vortex) in both the $R_1$ and $R_2$ cycles
of the torus, i.e. the Hopf texture. BOTTOM: $\Delta_1(x)$ has a
unit phase winding in only the $R_1$ cycle.
It is clear that there is a single level crossing in the TOP case,
meaning the ground state fermion parity is changed in the process.
The initial state has even fermion parity, so the final state, the
Hopf texture,must have odd fermion parity. The calculation is on the
cubic lattice model defined in the text, with $R_1=13$ and $R_2=5$
lattice units, and only points within the torus are retained.
Parameters used: $t=M=\lambda=1$, and pairing $|\Delta|=1$. }
\label{fig3}
\end{center}
\end{figure}

{\em Numerical Calculation:} We study numerically the microscopic
topological insulator model defined in reference \cite{hosur-2009}, with
a pair of orbitals ($\tau_z=\pm 1$) on each site of a cubic lattice.
The tight binding Hamiltonian $H = \sum_k[\psi^\dagger_k {\mathcal
H_k}\psi_k+H_{\rm pair}(k)]$, is written in momentum space using a
four component fermion operator $\psi_k$ with two orbital and two
spin components. Then, ${\mathcal H_k} = -2t \sum_{a=1}^3 \alpha_a
\sin k_a -m\beta_0 [\lambda +\sum_{a=1}^3 (\cos k_a -1)]$, where
$(\vec{\alpha},\,\beta_0)=(\tau_x\sigma_x,\,\tau_x\sigma_y,\,-\tau_y,\tau_z)$.
For $t,\,m>0$ a strong topological insulator is obtained when
$\lambda \in (0,2)$. In addition we introduce onsite singlet pairing
$H_{\rm pair} (k)= \Delta [\psi^\dagger_k]^{\rm T}\sigma_y
\psi^\dagger_{-k}+\,{\rm h.c. }$. Note, when $\lambda \approx
0,\,\vec{k}\approx (0,\,0,\,0)$, Eqn. \ref{HMF} is recovered as the
low energy theory.

The energy spectrum is studied as we interpolate between a
topologically trivial texture ($h=0$) and the Hopf texture ($h=1$).
We choose to define a torus shaped strong topological insulator with
trivial insulator (vacuum) on the outside, as in figure \ref{fig2}.
The surface is gapped by superconducting pairing $\Delta$, which in
the trivial texture is taken to be real $\Delta_0$. In the Hopf
texture, the superconducting pairing $\Delta_1$ has a phase that
winds around the surface, with a unit winding about both cycles of
the torus. This can be interpreted as 'vortices' inside the holes of
the torus. Note, the vortex cores are deep inside the insulators, so
there is a finite gap in the Hopf texture. We interpolate between
these two fully gapped phases be defining $\Delta(\lambda)=\lambda
\Delta_1 + (1-\lambda)\Delta_0$ and changing $\lambda=0 \rightarrow
1$. On the way, the gap must close since the two textures differ in
topology. We study the evolution of eigenvalues as shown in figure
\ref{fig3}. We find that exactly one pair of $\pm E$ eigenvalues are
pumped through zero energy. As argued below, this signals a change
in the fermion parity of the ground state on the two sides. Since
the trivial texture has even fermion parity from time reversal
symmetry, the Hopf texture must carry odd fermion number. No such
odd level crossings occur for topologically trivial textures (eg.
phase winding through only one cycle of the torus).

To see why the crossing of a $\pm E$ conjugate pair of levels
corresponds to a change in fermion number, consider a single site
model $H=E_0(c^\dagger c-cc^\dagger)$. This has a pair of single
particle levels at $\pm E_0$, which will cross if we tune $E_0$ from
say positive to negative values. However, writing this Hamiltonian
in terms of the number operators $H=E_0(2\hat{n}-1)$, shows that the
ground state fermion number changes from $n=0$ to $n=1$ in this
process. Thus the ground state fermion parity is changed whenever a
pair of conjugate levels cross zero energy.

{\em The Pfaffian:} Previously, the ground state fermion parity was
found by interpolating between two topological sectors. Can one
directly calculate the fermion parity for a given Hamiltonian's
ground state? We show this is achieved by calculating the Pfaffian
of the Hamiltonian in the Majorana basis. The Pfaffian of an
antisymmetric matrix is the square root of the determinant - but
with a fixed sign. It is convenient to recast the Hamiltonian in
terms of Majorana or real fermions defined via
$\psi_a=(\chi_{1a}+i\chi_{2a})/2$. Since a pair of Majorana fermions
anticommute $\{\chi_i,\,\chi_j\}=\delta_{ij}$, the Hamiltonian
written in these variables will take the form:
\begin{equation}
H = -i \sum_{ij} {\rm \bf h}_{ij}\chi_i \chi_j
\end{equation}
where ${\rm \bf h}_{ij}$ is an even dimensional antisymmetric
matrix, with real entries and the Majorana fields appear as a vector
$\chi = (\dots,\,\chi_{1a},\chi_{2a},\,\dots)$, where $a$ refers to
site, orbital and spin indices.
The $\pm E$ symmetry of the spectrum is an obvious consequence of
$h$ being an antisymmetric matrix. The ground state fermion parity
in this basis is determined via:
\begin{equation}
(-1)^F = sign \left[ {\rm Pfaffian}({\rm \bf h})\right]
\end{equation}

We numerically calculated the Pfaffian of a Hamiltonian with a
single Hopf texture for small systems and confirmed it has a
negative sign. In contrast, the trivial Hopf texture Hamiltonian has
positive Pfaffian in the same basis.

Finally, we mention that it is possible to confirm the numerical
results analytically, by solving for the low energy modes in the
vicinity of the vortex core, where the insulating mass term is set
to be near zero. The linking of vortices in the Hopf texture plays a crucial role in
deriving this result. In the Appendix, we discuss how this result is
connected to the three dimensional non-Abelions in the hedgehog
cores of Ref \onlinecite{PhysRevLett.104.046401}.

{\em A Two Dimensional Analog:} We briefly mention a two dimensional
analog of the physics described earlier. Note, eqn. \ref{HMF} with
the third component of momentum absent $p_z=0$, describes a quantum
spin Hall insulator (trivial insulator) when $m>0$ ($m<0$), in the
presence of singlet pairing $\Delta$. Again, as before a three
vector characterizes a fully gapped state, and the nontrivial
textures are called skyrmions ($\Pi_2(S_2)=Z$). A unit skyrmion can
be realized with a disc of quantum spin Hall insulator with
superconductivity on the edge, whose phase winds by $2\pi$ on
circling the disc. Again, one can show that the skyrmion charge $Q$
determines the fermion parity $(-1)^Q=(-1)^F$. An important
distinction from the three dimensional case is that the low energy
theory here has a conserved charge. If instead of superconductivity,
one gapped out the edge states with a time reversal symmetry
breaking perturbation which had a winding, then this charge is the
electrical charge. It is readily shown that the charge is locked to
the skyrmion charge $Q$. Hence odd strength skyrmions are
fermions\footnote{The effective field theory for the $\vec{n}$
vector in this case includes a Hopf
term\cite{PhysRevLett.51.2250,Abanov2000321}, which ensures
fermionic skyrmions}. This is closely analogous to the Quantum Hall
ferromagnet, where charged skyrmions also occur
\cite{PhysRevLett.64.1313,PhysRevB.47.16419}. Returning to the case
with pairing, since that occurs on a one dimensional edge, it is
difficult to draw a clear-cut separation between fermions and
collective bosonic coordinates, in contrast to the higher
dimensional version. Hence we focus on the 3DTIs.

\section{Effective Theory and Topological Term}
 The gap to the $\psi$ fermions never vanishes since
$|\vec{n}|>0$, so one can integrate out the fermions to obtain a low
energy theory written solely in terms of the bosonic order parameter
$\vec{n}$. How can this field theory describe a fermionic texture?
As described below, this is accomplished by a topological Berry
phase term which appears in the effective action for the $\hat{n}$
field.

In computing the topological term, it is sufficient to consider a
gap whose magnitude is constant $\vec{n}(r,t)=M\hat{n}(r,t)$.
Integrating out the fermion fields with action $S_f=\int
d^4\;x[\psi^\dagger
\partial_t \psi -H_f]$, (the integral is over space and (Euclidean)
time), one obtains the effective action for the bosonic fields:

\begin{equation}
 e^{-S_B(\hat{n}(r,t)} = \int {\mathcal D}\psi
{\mathcal D}\psi^\dagger e^{-S_f[\hat{n},\psi,\psi^\dagger]}
\end{equation}
This computation may be performed using a gradient expansion, i.e.
assuming slow variation of the $\hat{n}$ field over a scale set by
the gap. Two terms are obtained: $S_b = S_0+S_{\rm top}$. The first \footnote{While the
Dirac theory in \ref{HMF} has an O(3) symmetry, physically the
symmetry is lower, corresponding to O(2) charge rotations. Hence
other terms consistent with this lower symmetry are allowed in the
effective action, however, the topological term, which is the main
focus, does not depend on this detail.}
is a regular term that penalizes spatial variation:
$S_0=\frac1{2g}\int (\partial_\mu \hat{n})^2$.  The second is a
topological term which assigns a different amplitude to
topologically distinct spacetime configurations of $\hat{n}$.
Assuming $\hat{n}(\infty)=const$, these configurations are
characterized by a $Z_2$ distinction
($\Pi_4(S_2)=Z_2$)\cite{PatodiAtiyahSinger}. That is, there are two classes of
maps - the trivial map, which essentially corresponds to the uniform
configuration, and a non-trivial class of maps, which can all be
smoothly related to a single representative configuration
$\hat{n}_1(r,t)$. If the function $\Gamma[\{\hat{n}(r,t)\}]=0,\,1$
measures the topological class of a spacetime configuration, then
the general form of the topological term is $S_{\rm top} = i\theta
\Gamma[\{\hat{n}(r,t)\}]$. The topological angle $\theta$ can be
argued to take on only two possible values $0,\,\pi$, since
composing a pair of nontrivial maps, leads to the trivial map. Via
an explicit calculation, outlined below, we find $\theta=\pi$. Let
us first examine the consequences of such a term. The nontrivial
texture $\hat{n}_1(r,t)$ can be described as a Hopfion-antiHopfion
pair being created at time $t_1$, the Hopfion being rotated slowly
by $2\pi$, and then being combined back with the anitHopfion at a
later time $t_2$ \cite{PhysRevLett.51.2250}. The topological term assigns a
phase of $e^{i\pi}$ to this configuration. This is equivalent to
saying the Hopfion is a fermion, since it changes sign on $2\pi$
rotation.
%\footnote{This nontrivial spacetime texture may also be
%directly related to exchange of a pair of Hopfions.}

Calculating the topological term requires connecting the pair of
topologically distinct configurations. To do this in a smooth way
keeping the gap open at all times requires enlarging the order
parameter space for this purpose. If ${\bf m}(r,t,\lambda)$ is an
element of this enlarged state that smoothly interpolates between
the trivial configuration ${\bf m}(r,t,0)={\rm const.}$ and the
nontrivial one ${\bf m}(r,t,1)=\hat{n}_1(r,t)$, as we vary
$\lambda$, then one can analytically calculate the change in the
topological term $\partial S_{\rm top}/\partial \lambda$, and
integrate it to get the required result \cite{Elitzur1984205, Klinkhamer199141,
Abanov2000321,Abanov2000685}. The key technical point is finding a suitable enlargement
of our order parameter space $S_2$. Remembering that this can be
considered as $S_2=SU(2)/SO(2)$, we can make a natural
generalization $M_3 = SU(3)/SO(3)$. The latter has all the desirable
properties of an expanded space, eg. there are no nontrivial
spacetime configurations, so everything can be smoothly connected
($\Pi_4(M_3)=0$). This extension allows us to calculate the
topological term, (as explained in detail in the methods section),
which yields $\theta=\pi$.
% Unlike quantum Hall,
%one cannot simply measure the charge of the ground state to
%determine statistics. No conserved U(1). (i)Analytical calculation -
%linked vortices in the Dirac limit. Single zero mode pair crossing.
%(ii) Numerical calculation and Pfaffian calculation? (iii)
%Lagrangian calculation and the Berry's phase. (iv) Connection to 3D
%Majorana interpretation. (v) 2D QSHE case - skyrmion as fermion.

\section{Physical Consequences}
We now discuss two kinds of physical consequences arising from
fermionic Hopfions. The first relies on the dynamical nature of the
superconducting order parameter, while the second utilizes the
Josephson effect to isolate an anomalous response. The first class
conceptually parallels experiments used to identify skyrmions in
quantum hall ferromagnets. There, when skyrmions are the cheapest
charge excitations, they are detected on adding electrons to the
system \cite{PhysRevLett.74.5112}.

Consider surface superconductivity on a mesoscopic torus shaped
topological insulator as in Figure \ref{fig2}. The Hopf texture
corresponds to unit phase winding in each cycle of the torus. The
energy cost, $E_H = (\rho_s/2)\int (\nabla \phi)^2 d^2x$ is simply
proportional to the superfluid density $E_H=A \rho_s$, where $A$ is
in general an $O(1)$ constant. If we have $E_H < \Delta$ the
superconducting gap, then the Hopfion is the lowest energy fermionic
excitation. Tunneling a single electron onto the surface should then
spontaneously generate these phase windings in equilibrium.
Measuring the corresponding currents (eg. via an RF squid) can be
used to establish the presence of the Hopf texture. A daunting
aspect of this scheme is to obtain a fully gapped superconductor
with $\rho_s<\Delta$.

 \begin{figure}
\includegraphics[width=0.5\textwidth]{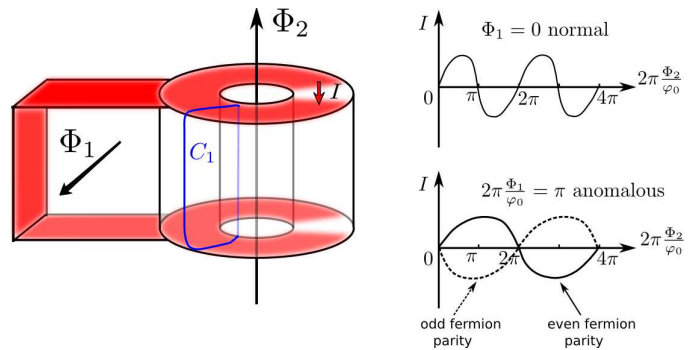}
\caption{Anomalous Josephson Response connected to fermionic
Hopfions: An annular cylinder of TI with pairing induced on the top
and bottom surfaces via proximity to a superconductor (present in
the red regions) is shown. Tuning the flux to $\Phi_1\approx
\varphi_0/2$, induces a vortex in the $C_1$ cycle. Now, the
Josephson effect on tuning $\Phi_2$ will be anomalous, with a part
that is not periodic in the flux $\varphi_0$. This is directly
related to the change in ground state fermion parity at
$\Phi_2=2\pi(\varphi_0/2\pi)$, where the Hopf texture is realized.
Adding a fermion inverts this current, indicating that the ground
state in this sector is at $\Phi_2 = 2\pi(\varphi_0/2\pi)$. If
however $\Phi_1\approx 0$, there is no vortivity in the $C_1$ cycle,
and the Josephson effect is the usual one, periodic in the flux
quantum $\varphi_0=h/2e$.}
 \label{fig4}
\end{figure}

A different approach relies on the Josephson effect as illustrated
in Figure \ref{fig4}. A hollow cylinder of topological insulators is
partially coated with superconductor on the top and bottom surfaces,
forming a pair of Josephson junctions. A unit vortex along the $C_1$
cycle can be induced by enforcing a phase difference of $\pi$
between the top and bottom surfaces, using the flux $\Phi_1=
{\mathcal \varphi}_0/2$ (where $\varphi_0=h/2e$ is the
superconductor flux quantum). Now, the vorticity enclosed by the
annulus determines the Hopf number, and hence the ground state
fermion parity. This vorticity can be traded for magnetic flux
$\Phi_2$ (parameterized via $f_2=2\pi (\Phi_2/\varphi_0)$) threading
the cylinder, since only $\nabla \phi -eA$ is gauge invariant.
Consider beginning in the ground state with $f_2=0$ and then tuning
to $f_2=2\pi$. One is now in an excited state since the ground state
at this point has odd fermion parity. This must be reflected in the
Josephson current $I$. We argue this implies doubling the flux
period of the Josephson current. Since $I=\partial
E(\Phi_2)/\partial \Phi_2$, the area under the $(I,\,\Phi_2)$ curve
:$\int_0^{\varphi_0}d\Phi_2 \, I[\Phi_2] = E[\varphi_0]-E[0]>0$ is
the excitation energy which does not vanish. Hence, the Josephson
current is not periodic in flux $\varphi_0$, as in usual Josephson
junctions. If we started with an odd fermion number to begin with,
then the state of affairs would be reversed - the ground state would
be achieved at multiples of $f_2=2\pi$. The ground state with a
particular fermion parity can be located by studying the slope of
the current vs phase curve. Since the energy of the ground state
increases on making a phase twist $\partial I/\partial \Phi_2 =
\partial^2 E/\partial \Phi_2^2>0$, it is associated with a positive
$I \,{\rm vs }\, \Phi_2$ slope. This positive slope will be at even
(odd) multiples of $f_2=2\pi$ for even (odd) fermion parity. If on
the other hand unit vorticity was not induced in the cycle $C_1$,
(eg. if $\Phi_1 \sim 0$), the Josephson relation would be the usual
one - i.e. one that is periodic in $f_2=2\pi$. This is summarized in
Figure \ref{fig4}.

Note, a similar anomalous $4\pi$ Josephson periodicity was pointed
out in the context pf the 2D QSH case with proximate
superconductivity in \cite{PhysRevB.79.161408}. We interpret this result in
terms of the fermionic nature of the solitons there - which lends a
unified perspective. In the 3D case, the Hopf texture allows one to
tune between the normal and anomalous Josephson effect by tuning the
$C_1$ vorticity via $\Phi_1$.

\section{Conclusions}
The low energy field theory of the superconductor-TI system was
derived and shown to possess a topological Berry phase term, which
leads to fermionic Hopf solitons. We note that topological terms are
particularly important in the presence of strong quantum
fluctuations. For example, in one dimension where fluctuations
dominate, the Berry phase term of the spin 1/2 Heisenberg
chain\cite{PhysRevLett.61.1029} leads to an algebraic phase. By
analogy, it would be very interesting to study the destruction of
superconductivity on a TI surface driven by quantum fluctuations.
The Berry phase, or relatedly, the fact that a conventional
insulator must break time reversal on the TI surface, will provide
an interesting  twist to the well know superconductor-insulator
transition studied on conventional substrates
\cite{PhysRevLett.62.2180}.

We acknowledge helpful discussions with C. Kane, A. Turner and S.
Ryu. A.V. and P.H. were supported by NSF DMR- 0645691.

\section{Appendix}
{\em A. Connections to 3D non-Abelian statistics}
\begin{figure}
 \includegraphics[width=0.48\textwidth]{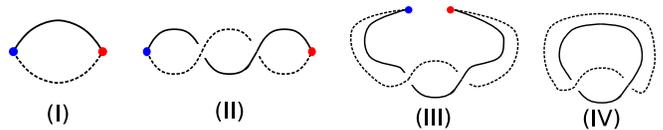}
\caption{(color online) Creating a Hopf soliton by a hedgehog-antihedgehog pair. The blue(red) dot is the hedgehog(antihedgehog). The solid and dotted lines are preimages of two different points on the 2-sphere. (I) creating a pair of hedge and antihedgehog - these could be a vortex-antivortex pair on the surface of a topological insulator (II) rotating the hedgehog by $2\pi$ while leaving the antihedgehog invariant (III)(IV) annihilating the hedgehog-antihedgehog pair. The final state in (IV) clearly shows linking number $1$ of the two preimage loops, which indicates the non-trivial Hopf index.}
\label{fig5}
\end{figure}

It is well known that vortices piercing a superconductor on the
topological insulator surface carry Majorana zero modes in their
cores \cite{fu:096407,PhysRevLett.104.046401}. In the $\vec{n}$
vector representation of Eq. (\ref{HMF}), this corresponds to a
hedgehog defect\cite{PhysRevLett.104.046401}, a singular
configuration where the vector points radially outwards from the
center. Although in this paper we only work with smooth textures, we
discuss below an indirect connection with those works. Note, we can
go from a trivial texture to the Hopf texture by creating a
hedgehog-antihedgehog pair, rotating one of them by an angle of
$2\pi$, and annihilating them to recover a smooth texture. This is
just the Hopf texture, as can be seen in the Figure \ref{fig5}.
However, as pointed out in Ref. \cite{PhysRevLett.104.046401}, in
the process of rotation, the Majorana mode changes sign. This
signals a change in fermion parity, consistent with our results.

{\em B. Calculation of Berry's phase}. We begin with:
\begin{align}
 H&=\mathbf\Psi^{\dagger}v_F(-i\partial_i\alpha_i)\mathbf\Psi+M\mathbf\Psi^{\dagger}(n_1\beta_0+n_2\beta_5\eta_x+n_3\beta_5\eta_z)\mathbf\Psi,\label{eq:fermion_model}
\end{align}
where, $\mathbf\Psi$ is the fermion in majorana basis (8-component),
$n_1$ is the insulator mass and $n_2,n_3$ are the superconductivity
masses, and the standard five anticommuting 4 by 4 Dirac matrices in
the Majorana basis (where the $\alpha$s are symmetric and $\beta$s
are antisymmetric matrices):
\begin{align}
\alpha_1=\sigma_z,\;\alpha_2=\tau_x\sigma_x,\;\alpha_3=-\tau_z\sigma_x,\;\beta_0=\tau_y\sigma_x,\;\beta_5=\sigma_y.
\end{align}
We further assume the order parameters $n_i$ are restricted to unit
2-sphere: $\sum_i n_i^2=1$ so that $\hat n=(n_1,n_2,n_3)$ is a unit
vector living on $S^2$.

We need to show that starting from this fermionic model
Eq.(\ref{eq:fermion_model}) and integrating out the fermions, the
obtained $S^2$-NLSM has an imaginary term (topological Berry's
phase) $i\theta H_{\pi_4(S^2)}(\hat n(x,y,z,t))$ with $\theta=\pi$
in the action (from now on we use $H_{\pi_n(M)}$ to denote the
homotopy index of a mapping.). Because this term is
non-perturbative, in order to compute it, we need to embed the
manifold $S^2$ into a larger manifold $M$ with $\pi_4(M)=0$, which
allows us to smoothly deform a $H_{\pi_4(S^2)}(\hat n(x,y,z,t))=1$
mapping to a constant mapping. This means that a
$H_{\pi_4(S^2)}(\hat n(x,y,z,t))=1$ mapping can be smoothly extended
over the 5-dimension disk: $ V(x,y,z,t,\rho):D^5\rightarrow M$
($\rho\in [0,1]$) such that on the boundary: $
V(x,y,z,t,\rho=1)=\hat n(x,y,z,t)$ and $ V(x,y,z,t,\rho=0)=V_0$ is
constant. With an extension $V$, we can perturbatively keep track of
the total change of Berry's phase when going from a constant mapping
to a non-trivial mapping.

How to find a suitable $M$? We note the global symmetry of model
Eq.(\ref{eq:fermion_model}) in the massless limit is
$U(1)_{chiral}\times SU(2)_{isospin}$, whose generators are:
\begin{align}
 U(1)_{chiral}:&\;\gamma_5=-i\beta_0\beta_5&SU(2)_{isospin}:&\;\eta_y,\gamma_5\{\eta_x,\eta_z\}.
\end{align}
In our convention, $\beta_0,\beta_5,\gamma_5$ are all anti-symmetric
matrices. Starting from a given mass, for instance,
$\mathbf\Psi^{\dagger}\beta_0\mathbf\Psi$, one can generate the full
order parameter manifold by action of $SU(2)_{isospin}$:
$\mathbf\Psi^{\dagger}W\beta_0W^{\dagger}\mathbf\Psi$.
$SU(2)_{isospin}$ is broken down to $U(1)$, the invariant group
generated by $\eta_y$. Thus the order parameter space is
$SU(2)/U(1)=S^2$.

Now we generalize the 8-component majorana fermion $\mathbf\Psi$ to
12-component $\mathbf{\tilde\Psi}$. The 2-dimensional
$\eta_{x,y,z}$-space is enlarged to a 3-dimensional space, and we
let the eight $\lambda_{i}$-matrices ($i=1,2...8$) of the standard
$SU(3)$ generators(see, e.g., page 61 in \cite{georgi}) act within
this 3-dimensional space, among which
$\lambda_2,\lambda_5,\lambda_7$ are antisymmetric while others are
symmetric. And $\lambda_{1,2,3}$ are just the old $\eta_{x,y,z}$
matrices. The symmetry of the generalized massless theory of
$\mathbf{\tilde\Psi}$:
$H=\mathbf{\tilde\Psi}^{\dagger}(-i\partial_i\alpha_i)\mathbf{\tilde\Psi}$
is $U(1)_{chiral}\times SU(3)_{isospin}$, where the generators of
the $SU(3)_{isospin}$ are
$\lambda_2,\lambda_5,\lambda_7$,$\gamma_5\{\lambda_1,\lambda_3,\lambda_4,\lambda_6,\lambda_8\}$.

Starting from a given mass
$\mathbf{\tilde\Psi}^{\dagger}\beta_0\mathbf{\tilde\Psi}$, we use
$SU(3)_{isospin}$ to generate the order parameter manifold:
$\mathbf{\tilde\Psi}^{\dagger}U\beta_0U^{\dagger}\mathbf{\tilde\Psi}\equiv
\mathbf{\tilde\Psi}^{\dagger}V\mathbf{\tilde\Psi}$, $U\in
SU(3)_{isospin}$. It is clear that the $SO(3)$ subgroup generated by
$\lambda_2,\lambda_5,\lambda_7$ is the invariant group and the order
parameter manifold is $M_3\equiv SU(3)/SO(3)$. We thus embed the
original order parameter manifold $S^2$ into $M_3$, and it is known
that $\pi_4(M_3)=0$(\cite{1751-8121-40-23-015}).

The idea is to smoothly extend a $H_{\pi_4(S^2))}=1$ mapping over
$D^5$ (denote by $V(x,y,z,t,\rho)$) by embedding $S^2$ into $M_3$.
In fact if we can extend a $H_{\pi_4(SU(2))}=1$ mapping over $D^5$
(denote by $U(x,y,z,t,\rho)$) by embedding $SU(2)_{isospin}$ into
$SU(3)_{isospin}$, it will generate the extension $V$ by $V=U\beta_0
U^{\dagger}$. This is because a $H_{\pi_4(S^2))}=1$ mapping can be
thought as a combination of a $H_{\pi_4(SU(2))}=1$ mapping and a
Hopf $H_{\pi_3(S^2)}=1$ mapping. Such an extension $U:D^5\rightarrow
SU(3)_{isospin}$ has already been explicitly given by Witten (see
Eq.(9-13) in Ref\cite{Witten1983433}), which we will term it
Witten's map. Witten's map was introduced to compute a $i\theta
H_{\pi_4(SU(2))}$ topological Berry's phase. Basically, on the
boundary $\partial D^5=S^4$ ($\rho=1$), Witten's map is a
$H_{\pi_4(SU(2))}=1$ mapping defined as a rotating
$H_{\pi_3(SU(2))}=1$ soliton (by $2\pi$) along the time direction.
This $H_{\pi_4(SU(2))}=1$ mapping at $\rho=1$ is smoothly deformed
into a trivial mapping at $\rho=0$ by embedding $SU(2)$ into
$SU(3)$.

We use Witten's map to generate $V$, with which we compute the
Berry's phase perturbatively by a large mass
expansion\cite{Abanov2000685,hosur-2009} of Lagrangian:
\begin{align}
 L=\mathbf{\tilde\Psi}^{\dagger}[\partial_{\tau}+i\partial_{i}\alpha_i]\mathbf{\tilde\Psi}+M\mathbf{\tilde\Psi}^{\dagger}V\mathbf{\tilde\Psi}\equiv \mathbf{\tilde\Psi}^{\dagger}[D]\mathbf{\tilde\Psi},
\end{align}
where $D=[\partial_{\tau}+i\partial_{i}\alpha_i+MV]$ and partition function is $Z=\int \mathcal{D}\mathbf{\tilde\Psi}^{\dagger}\mathcal{D}\mathbf{\tilde\Psi}\mathcal{D} V  e^{-\int d^{4}x L}$. After integrating out fermion, we obtain a NLSM of $V$: $\tilde L=-\frac{1}{2}\mbox{Tr}\ln[\partial_{\tau}+i\partial_{i}\alpha_i+MV]$. Here the factor $1/2$ is because we are integrating out majorana fields. If there is a variation $\delta V$, the variation of the imaginary part $\Gamma\equiv \mbox{Im}(\tilde L)$ is:
\begin{align}
 \delta\Gamma=\frac{-K}{2}\int dx^4\epsilon^{\alpha\beta\gamma\mu}\mbox{Tr}\{\gamma_5V\partial_{\alpha}V\partial_{\beta}V\partial_{\gamma}V\partial_{\mu}V\delta V\}
\end{align}
where $K=\int
\frac{d^4p}{(2\pi)^4}\frac{M^6}{(p^2+M^2)^5}=\frac{1}{192\pi^2}$.
Denoting $\partial_{\rho}=\partial_4$, after some algebra, the
Berry's phase can be written in the fully antisymmetric way:
\begin{align}
 \Gamma=\frac{-K}{10}\int dx^5\epsilon^{\alpha\beta\gamma\mu\nu}\mbox{Tr}\{\gamma_5V\partial_{\alpha}V\partial_{\beta}V\partial_{\gamma}V\partial_{\mu}V\partial_{\nu} V\}\label{eq:fully_asym}
\end{align}
In fact $\Gamma$ is not fully well defined because the ambiguity of
the extension of $\hat n(x,y,z,t)$ to the 5-disk: two different
extensions $V(x,y,z,t,\rho)$ can differ by a mapping $S^5\rightarrow
M_3$. We will soon show that this ambiguity means that $\Gamma$ is
well defined up to $\mbox{mod } 2\pi$.

As $V$ is generated by $U$, plugging $V=U\beta_0 U^{\dagger}, U\in SU(3)_{isospin}$ in Eq.(\ref{eq:fully_asym}), one can further simplify it by trace out the $\beta_{0,5}$ space. Firstly note that $\partial_{\mu}V=U[U^{\dagger}\partial_{\mu}U,\beta_0]U^{\dagger}$, and because $U^{\dagger}\partial_{\mu}U$ is an element of the $SU(3)_{isospin}$ Lie algebra spanned by $\lambda_2,\lambda_5,\lambda_7$,$\gamma_5\{\lambda_1,\lambda_3,\lambda_4,\lambda_6,\lambda_8\}$, $[U^{\dagger}\partial_{\mu}U,\beta_0]$ simply picks out the latter five generators. After some algebra, one finds
\begin{widetext}
\begin{align}
 \Gamma
=&\frac{1}{15\pi^2}\int dx^5\epsilon^{\alpha\beta\gamma\mu\nu}\mbox{Tr}(g^{-1}\partial_{\alpha}g)_{\perp}(g^{-1}\partial_{\beta}g)_{\perp}(g^{-1}\partial_{\gamma}g)_{\perp}(g^{-1}\partial_{\mu}g)_{\perp}(g^{-1}\partial_{\nu}g)_{\perp}\},\label{eq:g-form}
\end{align}
\end{widetext}
where $g$ is defined as the corresponding 3 by 3 $SU(3)$ matrix of
$U$: if $U$ is the exponential of
$a_1\lambda_2+a_2\lambda_5+a_3\lambda_7+a_4\gamma_5\lambda_1+a_5\gamma_5\lambda_3+a_6\gamma_5\lambda_4+a_7\gamma_5\lambda_6+a_8\gamma_5\lambda_8$,
then $g$ is the exponential of
$a_1\lambda_2+a_2\lambda_5+a_3\lambda_7+a_4\lambda_1+a_5\lambda_3+a_6\lambda_4+a_7\lambda_6+a_8\lambda_8$.
$g$ is nothing but the Witten's map. And
$(g^{-1}\partial_{\mu}g)_{\perp}$ denotes the symmetric part:
$(g^{-1}\partial_{\mu}g)_{\perp}=\big[(g^{-1}\partial_{\mu}g)+(g^{-1}\partial_{\mu}g)^T\big]/2$.

We simply need to compute $\Gamma$ by integration. Because it is
clear that $\Gamma$ can only be $0$ or $\pi$ (mod $2\pi$), a
numerical integration is enough to determine it unambiguously. We
performed integration Eq.(\ref{eq:g-form}) with Witten's map $g$ by
standard Monte Carlo approach, and find $\Gamma=(1.000\pm
0.005)\pi$. This proves that Hopf-skyrmion is a fermion. In addition
it confirms that $\Gamma$ is well-defined only up to mod $2\pi$:
different extensions of $g$ can differ by a doubled Witten's map is
known to have $H_{\pi_5(SU(3))}=1$, and the above calculation
indicate that this ambiguity only add an integer times $2\pi$ in
$\Gamma$.

\bibliographystyle{apsrev}
\bibliography{/home/ranying/downloads/reference/simplifiedying}
\end{document}